\documentclass[sigconf]{acmart}
\AtBeginDocument{%
  }

\setcopyright{acmlicensed}
\copyrightyear{2025}
\acmYear{2025}
\acmDOI{XXXXXXX.XXXXXXX}
\renewcommand\footnotetextcopyrightpermission[1]{}
\acmISBN{000-0-0000-0000-0/0000/00}

\usepackage[edges]{forest}
\usepackage[normalem]{ulem}
\usepackage{enumitem}
\usepackage{pifont}
\usepackage{multirow}
\usepackage{colortbl}
\usepackage{xcolor}
\usepackage{booktabs} % 引入该宏包用于更好的表格横线样式
\usepackage{scalefnt}
\usepackage{amsmath}
\usepackage{adjustbox}
\usepackage{makecell}
\usepackage{array}
\usepackage{threeparttable}

\begin{document}

%%
%% The "title" command has an optional parameter,
%% allowing the author to define a "short title" to be used in page headers.
\title{A Survey on Trustworthy LLM Agents: Threats and Countermeasures}

%%
%% The "author" command and its associated commands are used to define
%% the authors and their affiliations.
%% Of note is the shared affiliation of the first two authors, and the
%% "authornote" and "authornotemark" commands
%% used to denote shared contribution to the research.

\author{\textbf{Miao Yu$^{1,\dag}$, Fanci Meng$^{1,\dag}$, Xinyun Zhou$^4$, Shilong Wang$^1$, Junyuan Mao$^1$, Linsey Pang$^2$,\\ Tianlong Chen$^3$, Kun Wang$^4$, Xinfeng Li$^{4,*}$, Yongfeng Zhang$^5$, Bo An$^4$, Qingsong Wen$^{1,*}$}}
\affiliation{
$^1$Squirrel AI Learning, $^2$Salesforce, $^3$The University of North Carolina at Chapel Hill,\\ $^4$Nanyang Technological University, $^5$Rutgers University
\country{}}
% \affiliation{
% $^1$Squirrel AI Learning, $^2$Salesforce, $^3$Massachusetts Institute of Technology,\\ $^4$Nanyang Technological University, $^5$Rutgers University
% }

\thanks{Xinfeng Li and Qingsong Wen are the corresponding authors. $\dag$ denotes equal contributions. Contact: Miao Yu (fishthreewater@gmail.com), Xinfeng Li (lxfmakeit@gmail.com), and Qingsong Wen (qingsongedu@gmail.com)}

\newcommand{\xf}[1]{{{\textcolor{blue}{[Xinfeng: #1]}}}}
\newcommand{\checkicon}{\raisebox{-.25em}{\includegraphics[width=1em]{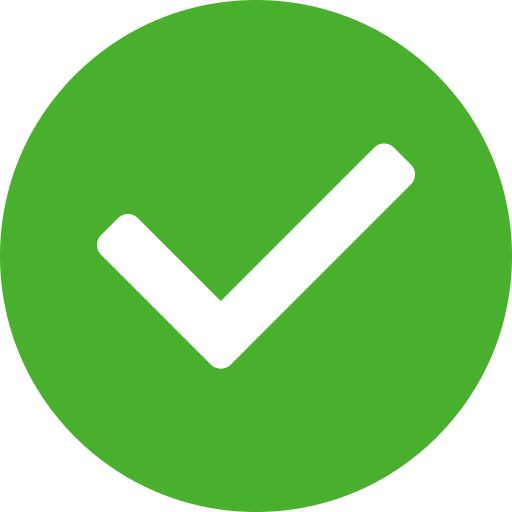}}}
\newcommand{\crossicon}{\raisebox{-.25em}{\includegraphics[width=1em]{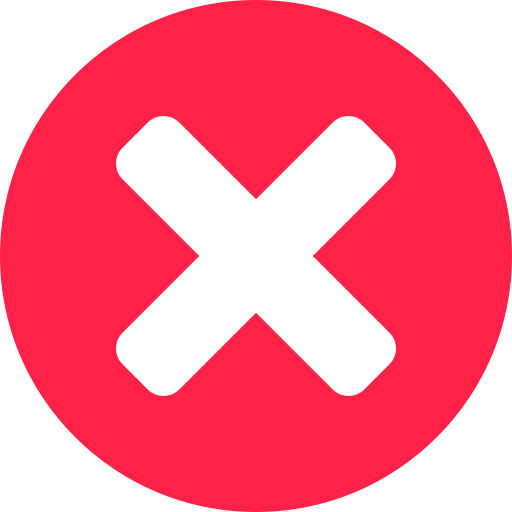}}}
%%
%% By default, the full list of authors will be used in the page
%% headers. Often, this list is too long, and will overlap
%% other information printed in the page headers. This command allows
%% the author to define a more concise list
%% of authors' names for this purpose.
\renewcommand{\shortauthors}{Yu et al.}

%%
%% The abstract is a short summary of the work to be presented in the
%% article.
\begin{abstract}
    With the rapid evolution of Large Language Models (LLMs), LLM-based agents and Multi-agent Systems (MAS) have significantly expanded the capabilities of LLM ecosystems. This evolution stems from empowering LLMs with additional modules such as memory, tools, environment, and even other agents. However, this advancement has also introduced more complex issues of trustworthiness, which previous research focusing solely on LLMs could not cover. In this survey, we propose the TrustAgent framework, a comprehensive study on the trustworthiness of agents, characterized by \textbf{modular taxonomy, multi-dimensional connotations, and technical implementation}. By thoroughly investigating and summarizing newly emerged attacks, defenses, and evaluation methods for agents and MAS, we extend the concept of Trustworthy LLM to the emerging paradigm of Trustworthy Agent. In TrustAgent, we begin by deconstructing and introducing various components of the Agent and MAS. Then, we categorize their trustworthiness into intrinsic (brain, memory, and tool) and extrinsic (user, agent, and environment) aspects. Subsequently, we delineate the multifaceted meanings of trustworthiness and elaborate on the implementation techniques of existing research related to these internal and external modules. Finally, we present our insights and outlook on this domain, aiming to provide guidance for future endeavors. For easy reference, we categorize all the studies mentioned in this survey according to our taxonomy, available at: \textcolor{blue}{\url{https://github.com/Ymm-cll/TrustAgent}}.
\end{abstract}

\maketitle

\section{Introduction}

The advent of large language models (LLMs) has catalyzed a paradigm shift in artificial intelligence systems \cite{zhao2023survey, minaee2024large, chang2024survey, wang2024survey}. The integration of LLMs as backbone with extra modules (e.g. memory \cite{zhang2024survey}, tool \cite{masterman2024landscape, qin2024tool}, and environment \cite{yang2024embodied}) as extensions produces the concept of ``LLM-based Agent'', transforming static neural network into dynamic cognitive subject capable of memory retrieval, tool utilization, and environmental interaction. Furthermore, the introduction of inter-agent communication has given rise to the more advanced concept of Multi-agent System (MAS), making the ``hyper LLM ecosystem'' become even more intricate, interactive and intelligent \cite{safegurad, zhang2025evoflow, masrouter, zhang2025multi, guo2024large, xie2024large}.

Extensive academic research and industry practices have validated this performance hierarchy: MAS > Single Agent > LLM \cite{li2023camel, wu2023autogen, talebirad2023multi, chan2023chateval}. However, the incorporation of additional modules is a double-edged sword, which has raised new concerns regarding trustworthiness across multiple dimensions, including safety, privacy, fairness, and truthfulness \cite{he2024emerged, wang2024large, xi2025rise}. From a risk perspective, introducing new modules expands the system's attack surface, potentially leading to unforeseen vulnerabilities \cite{yuan2024r, tang2024prioritizing, gan2024navigating}. On the other hand, this integration poses new challenges to existing defense mechanisms and trustworthiness evaluations, necessitating the expansion and upgrade of previous research that focuses solely on the trustworthiness of LLMs or a single agent \cite{zhang2024agent, hua2024trustagent}.

\begin{table}[t]
\centering
\setlength{\abovecaptionskip}{1pt}% 
\setlength{\belowcaptionskip}{1pt}%
\caption{Comparison between TrustAgent and other surveys.}
\label{comparison}
\begin{adjustbox}{width=1\linewidth}
\begin{tabular}{c|c|cccccc}
    \hline
    \rowcolor{gray!30} \textbf{Survey} & \textbf{Object} & \textbf{\makecell{Multi-\\Dimension}} & \textbf{Modular} & \textbf{Technique$^\ddagger$} & \textbf{MAS}$^\star$\\
    \hline
    Liu et al. \cite{liu2023trustworthy} & LLM & \checkicon & \crossicon & Atk/Eval & \crossicon\\
    \rowcolor{gray!10} Huang et al. \cite{huang2024trustllm} & LLM & \checkicon & \crossicon & Eval & \crossicon\\
    \hline
    He et al. \cite{he2024emerged} & Agent & \crossicon & \crossicon & Atk/Def & \crossicon\\
    \rowcolor{gray!10}Li et al. \cite{li2024personal} & Agent & \checkicon & \crossicon & Atk & \crossicon\\
    Wang et al. \cite{wang2024large} & Agent & \crossicon & \crossicon & Atk & \crossicon\\
    \rowcolor{gray!10}Deng et al. \cite{deng2024ai} & Agent & \crossicon & \checkicon & Atk/Def & \checkicon\\
    Gan et al. \cite{gan2024navigating}  & Agent & \checkicon & \crossicon & Atk/Def/Eval & \crossicon\\
    \hline
    \rowcolor{gray!10} \textbf{TrustAgent (Ours)} & LLM + Agent & \checkicon & \checkicon & Atk/Def/Eval & \checkicon\\
    \hline
\end{tabular}
\end{adjustbox}
\begin{tablenotes}[flushleft]
    \item[] \hspace{-2pt}\scriptsize 
    $\ddagger$: Attack (Atk), Defense (Def), and Evaluation (Eval) denote TrustAgent's broad technique view.\\
    $\star$: Multi-agent System (MAS).
    % \item[] \vspace{-2pt}\hspace{-2pt}\small 
\end{tablenotes}
\vspace{-1em}
\end{table}

\begin{figure*}[t]
    \centering
    \includegraphics[width=\linewidth]{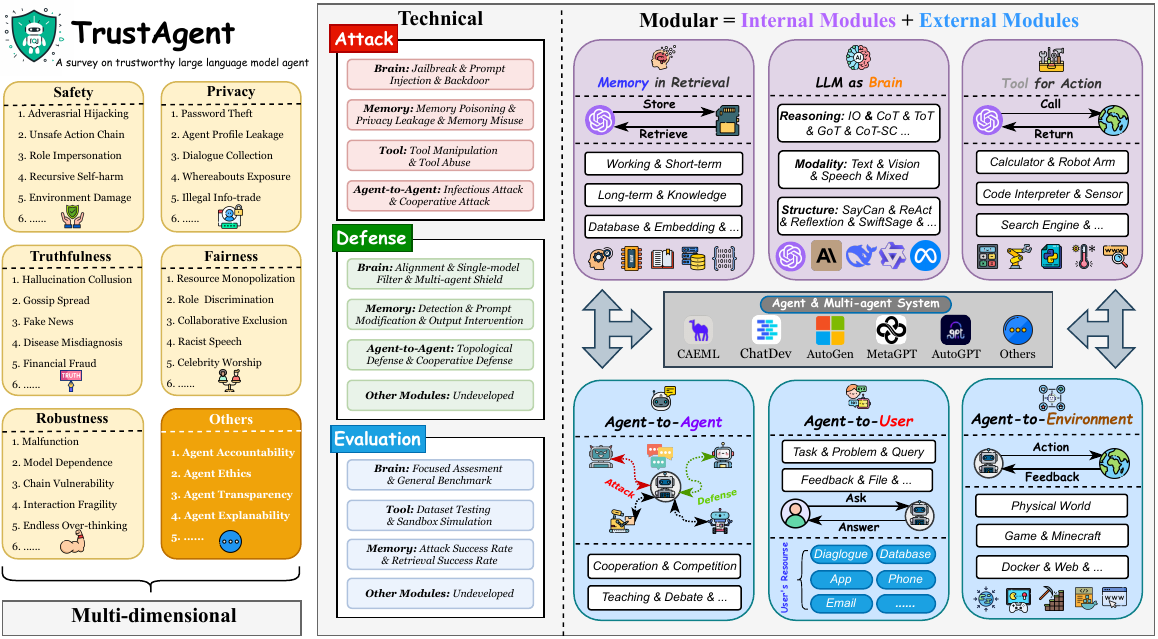}
    \vspace{-2em}
    \caption{Overview of our TrustAgent taxonomy, featuring multi-dimensional (Left), technical (Middle), and modular (Right).}
    \label{fig:trustagent intro}
    \vspace{-1em}
\end{figure*}

% 画个表格对比一下
Previous survey efforts have delved deeply into the realm of trustworthy LLMs. Liu et al. \cite{liu2023trustworthy} dissect trustworthiness into seven major categories, focusing specially on establishing standards and guidelines for LLM alignment. Similarly, Huang et al. \cite{huang2024trustllm} interpret trustworthiness from 6 perspectives but concentrate on creating benchmarks to evaluate trustworthiness. However, these studies are only partially effective in agent scenarios, highlighting an urgent need to address new trustworthiness issues arising from the introduction of additional modules. Other specialized surveys on trustworthiness in agents primarily focus on the sub-aspects like security and privacy and largely overlap with the content of Trustworthy LLMs \cite{wang2024large, wang2024large, gan2024navigating}. In fact, some works merely address the new issues arising from LLMs serving as the ``brain'' module of agents \cite{he2024emerged, li2024personal}, while overlooking the unexplored challenges introduced by other additional modules. To highlight our innovations, we compare TrustAgent with other surveys in Table \ref{comparison}.
% \xf{We should draft a comparison table first, and then maybe rewrite this paragraph to highlight differences between our work with prior ones; At least, "Our comparison between this survey to previous studies across xxx dimensions is shown in Table. 1."}

% 我们的方法论
To this end, we propose the TrustAgent framework, as illustrated in Figure \ref{fig:trustagent intro}, extending the research realm of previous trustworthiness surveys to the new context of agents and MAS. Our taxonomy in TrustAgent exhibits the following features: \textbf{(I) Modular.} TrustAgent rigorously categorizes trustworthiness issues based on the internal and external components of agents, specifically divided into intrinsic and extrinsic aspects. The former includes the trustworthiness of the brain, memory, and tools, while the latter encompasses the parts related to users, other agents, and the environment. 
% \xf{We may highlight the latter is caused by the ``potential interaction threats''.} 
\textbf{(II) Technical.} TrustAgent focuses on the implementation of trustworthy agents, providing a comprehensive summary and outlook on the relevant technology stack from three aspects: attack, defense, and evaluation (a comprehensive picture in Appendix \ref{taxonomy}). \textbf{(III) Multi-dimensional.} TrustAgent expands the dimensions of LLM trustworthiness to the context of both single agent and MAS, specifically categorized into: safety, privacy, truthfulness, fairness and robustness (with specific definitions in Appendix \ref{definition}), by involving existing works in all these dimensions.

% 内生

% 外生

% 类比于人类，我们将agent定义为一个独立的智能体，其本身即内部模块为大脑，记忆以及工具三部分。具体来说，brain模块即是执行推理、决策、反思等任务的LLM中心，不同于以往将LLM当作即时性聊天助手等用途，作为agent大脑的LLM在其中扮演的是持续性的进行任务求解、环境感知等任务的中枢。记忆模块则可以简单有效的分为工作记忆和长期记忆，其中工作记忆是该agent在执行某任务时行为、思维的历史记录，主要形式为文字日志等，而长期记忆则是类似于人类积累的各种先验知识，具体可以由文本知识库，向量数据库等并结合RAG技术来检索。工具则是agent与外界进行交互的核心，既充当从外接收集信息的媒介，又负责将内部决策出的行为反映到外界中。具体实现形式可以是api函数、感知器、具身机器人等。

% 除了agent内部的三模块外，我们将与agent进行交互的外界模块分为用户、其他agent和环境。

In each subsection, we begin with an overview of the current module's mechanisms and roles within the agent system, then explore its trustworthiness issues from well-categorized perspectives such as attacks, defenses, and evaluations. Finally, we provide inspiring insights and outline potential future research avenues.

In summary, our contributions can be listed below:
\begin{itemize}[leftmargin=*]
    \item \textbf{Thorough and Latest Survey.} We present a thorough and contemporary analysis of the trustworthiness in LLM-based agent systems, covering a wide spectrum of architectures including single LLM, individual agent, and MAS framework.
    \item \textbf{New Technique-oriented Taxonomy.} Our taxonomy centers on techniques for compromising, achieving, and evaluating trustworthiness, updating old paradigms to the agent context and outlining new technical paradigms within the agent framework.
    \item \textbf{Insightful Future Directions.} For each module's trustworthiness, we identify current vulnerabilities and outline future directions, urging researchers to delve deeper into this area.
\end{itemize}

% \textbf{Roadmap.} In this survey, we aim to provide a comprehensive overview of recent studies on Agent and MAS trustworthiness, covering attacks, defenses, and evaluations related to each module. Section \ref{Intrinsic} provides a thorough examination of the intrinsic components of LLM agents and the related trustworthiness challenges. We offer a detailed analysis of the auxiliary modules, with Section \ref{brain} focusing on the "brain" (LLM) module, Section \ref{memory} exploring the memory module, and Section \ref{tool} discussing the tool module. In Section \ref{Extrinsic}, we investigate the extrinsic interactions of agents and their associated trustworthiness challenges, analyzing agent-to-agent interactions in Section \ref{agent}, interactions within diverse environments in Section \ref{environment}, and agent-user interactions in Section \ref{user}. Finally, Section \ref{conclusion} presents a comprehensive conclusion.

% \xf{Overall, the introduction is good. Perhaps we can follow the style of \textbf{Safety at Scale} (https://arxiv.org/pdf/2502.05206), which outlines its key contributions within the introduction. First, we should ensure the main body is well-organized and polished. Then, each section can summarize its core contributions, creating a clear and compelling introduction that effectively highlights our work and persuades reviewers.}

% \xf{We should plot an overall TrustAgent framework figure to demonstrate how we organize intrinsic and extrinsic trustworthiness}

\section{Intrinsic Trustworthiness} \label{Intrinsic}
In this section, we focus on the trustworthiness of the internal modules of agent systems. In our definition, the agent system is an independent entity with human-like cognition, composed of brain (\textbf{Sec. \ref{brain}}) with memory (\textbf{Sec. \ref{memory}}), and behavior in the tool form (\textbf{Sec. \ref{tool}}). Due to the different functions and natures of these modules, the resulting trustworthiness issues vary. To provide in-depth analysis and instructive insight, we first introduce the role and functions of each module in the agent system and then summarize technical paradigms for attack, defense, and evaluation methods.

\subsection{Trustworthy LLM as Brain} \label{brain}
LLM agents consist of a central LLM ``brain'' module  \cite{wang2023describe}. The brain serves as the core reasoning and decision-making center, integrating inputs from various auxiliary modules to guide the agent towards its goal. Though the powerful brain connected with other modules enhances the agent's ability, this integration expands the potential attack surface, making the system more vulnerable to trustworthiness threats. In this section, we center on brain-related attacks, defenses, and evaluations, with illustration in Figure \ref{fig:brain}.
% According to the manipulation mechanisms (as illustrated in Figure~\ref{fig:brain}), we categorize attacks into three classes:

\subsubsection{Attacks.} 
\begin{figure}[t]
    \centering
    \includegraphics[width=1\linewidth]{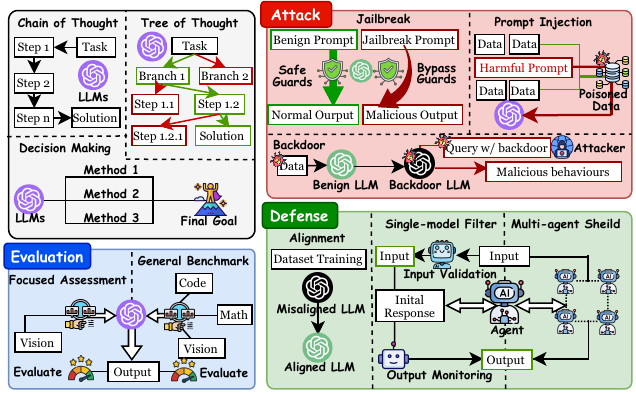}
    \vspace{-2.5em}
    \caption{The framework of agent brain's working mechanisms and its attack-defense-evaluation paradigm.}
    \label{fig:brain}
    \vspace{-2em}
\end{figure}

Unlike single-LLM systems, the brain module in agent or MAS experiences more frequent and complex information dynamics. The textual and visual inputs from the internal and external provide attackers with more interfaces and methods to compromise the trustworthiness of the core brain module. We do not discuss Misalignment attacks (e.g., fine-tuning an LLM to break its alignment) on agents here, as attackers typically lack access to modify the internal parameters of the agents. According to the manipulation mechanisms (as illustrated in Figure~\ref{fig:brain}), we categorize attacks into three paradigms:

% 我们没有在这里讨论agent的misalignment攻击，因为攻击者通常对agent内部参数没有查改权限。

\textbf{Jailbreak} attempts to bypass the aligned trustworthy mechanisms within the agent brain via human-designed or optimized adversarial prompts \cite{kumar2023certifying}. Early methods, such as GCG and its variants \cite{zou2023universal,jia2024improved,wang2024attngcg}, optimize adversarial suffixes to manipulate responses, coercing a ``Sure'' reply to malicious queries. MRJ-Agent \cite{wang2024mrj} designs a single attack agent to automatically generate covert jailbreak prompts, enhanced by information from multi-round dialogue. PrivAgent \cite{nie2024privagent} trains a single LLM as an attack agent through reinforcement learning to generate jailbreak prompts that can induce the target model to leak system prompts or training data. Furthermore, by leveraging multi-agent collaboration, Evil Geniuses \cite{tian2023evil} and PANDORA \cite{chen2024pandora} construct MAS as attack optimizers to perform role-specific jailbreak on the brain of target agents or MAS. They reinforced jailbreak effectiveness through Red-Blue exercises and MAS multi-step reasoning, respectively. Additionally, interactions between different brains give rise to viral jailbreak. Specifically, AgentSmith \cite{gu2024agent} and Tan et al. \cite{tan2024wolf} optimize self-replicating images to attack the brain of a single agent and discover an exponential spread of infectious jailbreak at MAS-level.

\textbf{Prompt Injection} embeds malicious prompts to override the original instructions to the agent brain, thereby manipulating its output or behavior \cite{liu2023prompt, perez2022ignore}. Early research manually crafts instructions and injects them into data which may be retrieved during inference, potentially causing the agent brain to deviate from intended tasks \cite{greshake2023not}. Later studies automate this process via gradient-based or optimization-based methods \cite{liu2024automatic, shi2024optimization}. Beyond injecting prompts in text modality, Bagdasaryan et al. \cite{bagdasaryan2023abusing} explore that multi-modal adversarial perturbations embedded in images or audio can manipulate the agent's brain to follow the attacker's instructions. The interaction between the agent's brain and other modules also provides more surfaces for prompt injection. Zhang et al. \cite{zhang2024breaking} exploit various interactions across multiple surfaces to mislead the agent's brain into performing repetitive or irrelevant actions, inducing logical errors and causing malfunctions.

 % 可能之后要加一块内容写data extraction 先放着。
%    Under \textit{Information Disclosure Attacks}, self-prompt calibration~\cite{zhang2024membership} shows the practicality of determining whether specific data were part of the training set. Nasr et al. reveal that through a specific prompt strategy, the model is induced to leak training data~\cite{nasr2023scalable}. PLLeak~\cite{hui2024pleak} enhances adversarial queries to coax system prompts from the target model, while other research reconstructs user prompts by analyzing input-output patterns~\cite{sha2024prompt,yang2024prsa}. Prompt Infection~\cite{lee2024prompt} exploits malicious prompts to spread across multi-agent systems, enabling data theft and misinformation.

\textbf{Backdoor} attacks involve the insertion of malicious triggers during training, which will be retrieved during inference, forcing predefined generation in the agent's brain for specific inputs \cite{zhou2025survey}. Yang et al. \cite{yang2024watch} categorize agent backdoor attacks into two main types: The first type manipulates the final output distribution, by inserting triggers in different phases (query, observation), while the second one inserts triggers in agent's thinking process to introduce malicious intermediate reasoning without altering the final output. The first type is exemplified by DemonAgent \cite{zhu2025demonagent}, which fragments the original backdoor into multiple sub-backdoor segments and employs dynamic encryption, making the backdoor continuously change during execution. BLAST \cite{yu2025blast} follows the second type and enables ``infectious backdoor'', implanting a backdoor in a single agent and using it to influence other agents' reasoning.

\subsubsection{Defenses.} The agent brain is not only more complex than a single LLM in terms of interactions, but also faces more types of attacks. Therefore, designing corresponding defenses to ensure its trustworthiness across all dimensions deserves holistic study. Drawing on the defense mechanisms' operational scopes, we organize defense approaches into three paradigms:

% Unlike single-LLM systems, the frequent and complex information dynamics within the brain module of an agent, as well as the collaborative interactions among the brain modules of multiple agents, also pose higher demands for effective defense.

% The frequent and complex information dynamics in the brain module of an agent system，以及多个agent的大脑模块之间的协同交互也对有效防御提出了更高要求

\textbf{Alignment} ensures that LLM agents operate in accordance with human values and ethical principles through fine-tuning, updating reward functions, etc. For instance, some studies embed human values directly into agents through intrinsic reward functions \cite{tennant2024moral} or neuro-symbolic rule-learning frameworks \cite{frisch2024llm}, validated in open-world environments like Minecraft. Another work fine-tunes agents with the data generated in social simulation scenarios \cite{pang2024self} to align agents with nuanced human behaviors and beliefs. Additionally, iterative alignment through user feedback and belief networks is leveraged to personalize agent behavior and reduce reliance on expert oversight, such as using historical edits \cite{gao2024aligning} or empirically derived human belief structures \cite{zhang2024large}.

% \textbf{Alignment.} Alignment involves modifying the model's training process or inference strategies to enhance its resilience.
% Reinforcement Learning from Human Feedback and its variants~\cite{ouyang2022training,bai2022training,xiong2023iterative,dong2024rlhf} aligns LLMs by iterative optimization against human preference-based reward models. Later researches combine reinforcement learning with other technologies. PE-VDN~\cite{gohari2023privacy} integrates RLHF into multi-agent learning environments to ensure safe interaction with external data. While Tennant et al. \cite{tennant2024moral} directly encode moral objectives into intrinsic reward functions and fine-tune LLM agents through Reinforcement Learning. And in order to allow the user to customize reward functions, Decoding-time Alignment techniques \cite{huang2024deal} further improve safety by adjusting responses dynamically at generation time.

\textbf{Single-model Filter} uses an external model to prevent attacks through input and output monitoring on the agent brain. Ayub et al. \cite{ayub2024embedding} employ traditional embedding classifiers to detect prompt injection by analyzing input semantic features, while Kwon et al. \cite{kwon2024slm} use a small model for harmful query detection. As for LLM filter, StruQ \cite{chen2024struq} mitigates prompt injection by converting inputs into structured queries with a specially trained LLM. ShieldLM \cite{zhang2024shieldlm} fine-tunes an LLM safety detector that aligns with safety standards. Additionally, agent serves as an even more powerful filter. For example, GuardAgent \cite{xiang2024guardagent} and AgentGuard \cite{chen2025agentguard} utilize a guardrail agent to protect target agents via safety constraint generation, action checking, and tool-use validation.

% Recent studies employ small models to discriminate malicious prompts \cite{ayub2024embedding, kwon2024slm}

% 其他工作使用embedding-based classifiers to detect prompt injection attacks by analyzing the semantic features of the input
% GuardAgent uses a guardrail agent to protect target agents by dynamically checking whether their actions comply with specified safety guard requirements.
% AgentGuard leverages the inherent capabilities of the LLM orchestrator to autonomously identifies and validates unsafe tool-use workflows and generates safety constraints to restrict the agent's behavior.

\textbf{Multi-agent Shield} leverages collaborative MAS to enhance the trustworthiness of the target agent's brain. The main difference of these guard MAS lies in their communication pattern. Specifically, one pattern is multi-agent debate, where agents critique the reasoning processes of other components of the brain over multiple rounds to reach a consensus final answer \cite{du2023improving}. Apart from debating, Kwartler et al. \cite{kwartler2024good} use a reviewing agent to provide feedback and correct the target agent's generation. AutoDefense \cite{zeng2024autodefense} assigns different roles to each agent, allowing smaller models to collaborate and protect the brain of the targeted agent from jailbreak attacks.

\subsubsection{Evaluation.} Different from static LLMs, agents continuously adapt to their environment, making their behavior highly context-dependent and variable, which also complicates the evaluation. Current evaluation methodologies for agent brain trustworthiness can be categorized along two dimensions:

\textbf{Focused Assessment} assesses the trustworthiness of agent brain when facing certain types of attacks or in specific domains. InjecAgent \cite{zhan2024injecagent} and AgentDojo \cite{debenedetti2025agentdojo} benchmark the brain's vulnerability to \textit{Indirect Prompt Injection}, while DemonAgent \cite{zhu2025demonagent} introduces AgentBackdoorEval for \textit{Backdoor} scenarios. RedAgent \cite{xu2024redagent} addresses context-aware red teaming, enabling test cases of context-sensitive \textit{Jailbreak}. In addition to evaluating brain trustworthiness in specific domains, RiskAwareBench \cite{zhu2024riskawarebench} introduces an automated framework for assessing \textit{physical risk awareness} in LLM-based embodied agents. RedCode \cite{guo2025redcode} provides a benchmark for evaluating the \textit{code} agents under risky code execution and generation with challenging test cases.

\textbf{General Benchmark} literally involves systematic evaluations to test the agent brain in diverse domains and from multiple dimensions. S-Eval \cite{yuan2024s} automates multiple dimensional and open-ended safety evaluation for LLMs with four hierarchical levels consisting of eight risk dimensions. While BELLS \cite{dorn2024bells} evaluates LLM Safeguard with a structured collection of tests, including established failure tests, emerging failure tests, and next-gen architecture tests. Meanwhile, recent efforts have introduced comprehensive benchmarks for LLM agents. Agent-SafetyBench \cite{zhang2024safetybench} and Agent Security Bench \cite{zhang2024asb} are benchmarks that evaluate LLM agents with multiple scenarios and different types of attack/defense methods. AgentHarm \cite{andriushchenko2024agentharm} contains 110 explicitly malicious agent tasks to evaluate LLM agent misuse, covering multiple harmful categories, including fraud, cybercrime, and harassment. RJudge \cite{yuan2024r} benchmarks the trustworthiness of multi-turn interactions based on agent interaction records, with 27 risk scenarios across multiple application categories and risk types.

\subsubsection{Insight.} Current collaborative attacks in MAS can spread from a single compromised agent's brain to multiple agents' brain modules \cite{gu2024agent, yu2025blast}. Therefore, developing collaborative security mechanisms that enable agents to monitor and validate each other's actions is crucial. For example, a distributed consensus protocol can be implemented, where agents collectively verify and agree on critical decisions before execution. Additionally, current evaluations primarily rely on static datasets. However, given the agent brain module's frequently dynamic interactions with external information, such evaluations are clearly insufficient. To address this, dynamic evaluation mechanisms should be implemented to better simulate real-world scenarios. For example, a continuous learning-based evaluation framework can be designed, where agents are tested in real-time environments with evolving data streams.

% although 现在有很多工作对llm的alignment进行研究，但是agent的alignment工作还很少，agent由于自身特性，即使目标与人类价值观对齐，agent的capabilities也可能与人类意图相悖，因此对agent的misalignment研究十分必要

% Although there is considerable research on the alignment of LLMs, there is still limited work on agent alignment. Due to the inherent characteristics of agents, even if their goals align with human values, their capabilities may still conflict with human intentions. Therefore, research on agent misalignment is crucial

\subsection{Trustworthy Memory in Retrieval} \label{memory}
\begin{figure}[t]
    \centering
    \includegraphics[width=1\linewidth]{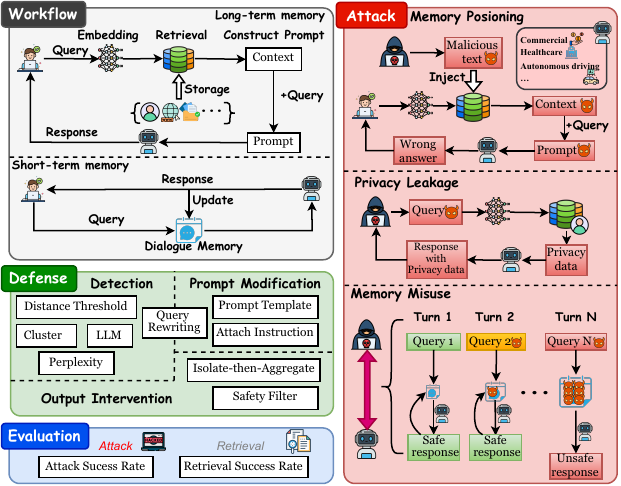}
    \vspace{-2.5em}
    \caption{The framework of the agent's memory utilization workflow and its attack-defense-evaluation paradigm.}
    \label{fig:memory}
    \vspace{-2em}
\end{figure}

Memory mechanisms in agent systems are crucial for interaction with the environment and user, categorized into long-term and short-term memory. Long-term memory is often associated with RAG, utilizing vector database to store real-world data for generation tasks, while short-term memory holds real-time interaction history, such as dialogue context or task logs. These mechanisms enhance agent capabilities but also introduce trustworthiness risks. In this section, we analyze the trustworthiness challenges posed by memory, focusing on attacks, defenses, and evaluation strategies. The overall framework is shown in Figure \ref{fig:memory}.

\subsubsection{Attack.}
We categorize attacks related to memory into three types: Memory Poisoning, Privacy Leakage, and Memory Misuse.

\textbf{Memory Poisoning} refers to attackers injecting malicious data into the \textit{long-term memory} \cite{xiang2024certifiably, chen2025agentpoison, zou2024poisonedrag, zhong2023poisoning, zhang2024agent, gu2024agent}, which will be retrieved and then mislead the agent system to generate incorrect outputs, undermining its truthfulness. This attack's danger stems from its stealthiness: once malicious data is injected into memory, it may continuously influence the agent until detected and removed. Specifically, we summarize the following attack paradigms: \ding{182} \textit{\textbf{Memory Injection.}} PoisonedRAG \cite{zou2024poisonedrag} optimizes malicious text through retrieval and generation conditions, enabling this text to be easily retrieved, while GARAG \cite{cho2024typos} continuously uses genetic algorithms to optimize adversarial examples for injection. Additionally, RobustRag \cite{xiang2024certifiably} and Zhong et al. \cite{zhong2023poisoning} demonstrate that injecting even a small amount of malicious information into vector database can successfully attack agents with high probability. \ding{183} \textit{\textbf{Backdoor.}} AgentPoison \cite{chen2025agentpoison} optimizes backdoor triggers and attaching them to queries, enhancing retrieval possibility of malicious samples.

% Obviously, backdoor make the malicious text in the database easier to be retrieved, reducing the robustness of the agent system.
% Once malicious texts are injected into the external database, user queries may retrieve data that is harmful to the task, thereby making it difficult for the agent to generate accurate results.
% Malicious text injection primarily focuses on creating text that can be retrieved and has the potential to mislead agents, with the differences among various attacks lying precisely in this aspect.

\textbf{Privacy Leakage} refers to attackers leveraging the connection between agent and its \textit{long-term memory} to steal the stored private data \cite{li2025commercial, zeng2024good, anderson2024my}. Beyond data theft, it enables criminal activities like phishing and identity trafficking, dramatically increasing privacy risks. Technically, we outline the following attack methods: \ding{182} \textit{\textbf{Jailbreak.}} RAG-Thief \cite{jiang2024rag} continuously optimizes anchor queries and adversarial commands, forming jailbreak prompts to conduct data stealing. RAG-MIA \cite{anderson2024my} designs specific jailbreak templates to extract private data from vector database. \ding{183} \textit{\textbf{Embedding Inversion}} directly restores the original data from its embedding. Specifically, Morris et al. \cite{morris2023text} iteratively optimize the text hypothesis to bring its embedding closer to the target embedding, while Li et al.\cite{li2023sentence} reconstruct the data via a generative decoder.

% It is crucial to emphasize that privacy leakage extends beyond data theft; it can enable severe criminal activities like phishing and identity theft, significantly escalating security risks for users.

% By re-embedding and refining the hypothetical text, gradually enhance the accuracy of restoration. 

% Zeng et al. \cite{zeng2024good} and Wang et al. \cite{wang2025unveiling} has shown that the use of RAG and memory systems increases the risk of privacy breaches. 

% Li et al.\cite{li2023sentence} propose generative embedding inversion attack (GEIA), which, based on a black-box setup, leverages a powerful generative decoder to utilize sentence embeddings as initial token representations, training the model to generate sequences that closely resemble the context of the original sentences.

% It is important to note that the consequences of privacy leakage are not limited to data theft; they can also be used to carry out more severe criminal activities such as phishing attacks and identity theft, posing greater security threats to users.

\textbf{Memory Misuse} crafts specific query sequences to \textit{gradually} bypass intrinsic safety alignment via multi-turn interaction, leveraging the storage property of agent \textit{short-term memory} \cite{russinovich2024great, li2024llm, cheng2024leveraging, priyanshu2024fractured, agarwal2024prompt, tong2024securing}. Therefore, the safety of an agent system is reduced. From a technical perspective, we outline the following two attack methodologies: \ding{182} \textit{\textbf{Jailbreak.}} Agarwal et al. \cite{agarwal2024prompt} send an initial query along with an attack prompt in the first round, followed by a flattering challenge and a reiteration of the attack prompt in the second round, while Russinovich et al. \cite{russinovich2024great} begin with harmless questions related to the target task and gradually steer harmful generation. \ding{183} \textit{\textbf{Backdoor.}} Tong et al.\cite{tong2024securing} utilize multi-turn dialogue memory to conceal backdoor triggers, activating malicious responses only when all triggers appear, thus stealthily achieving the attack.
As a core component of agents, defense technologies against short-term memory misuse are even more important.

% Short-term memory is a widely used and common component in agents. Therefore, developing effective defense techniques to prevent its misuse is a critical and essential research issue.

% , exploiting agent's pattern-following behavior and focusing on recent context to bypass safety alignment mechanisms.

% \textbf{Short-term memory} enhances performance by enabling the agent to progressively understand user intentions or interact with the environment, yet simultaneously introduces security vulnerabilities \cite{russinovich2024great, li2024llm, cheng2024leveraging, priyanshu2024fractured, agarwal2024prompt, tong2024securing}. Attackers may circumvent the agent's security measures through carefully designed dialogue sequences to achieve jailbreak or other malicious objectives.

% \textbf{Short-term memory} enables the agent to progressively understand user intentions and accomplish tasks, thereby enhancing its performance. However, this also provides opportunities for attackers. Attackers may exploit this incremental interaction mechanism by designing specific dialogue sequences or conceal actual attacks within continuous interactions, thus circumventing the agent's security protections to achieve jailbreaks or other malicious objectives \cite{russinovich2024great, li2024llm, cheng2024leveraging, priyanshu2024fractured, agarwal2024prompt, tong2024securing}.

\subsubsection{Defense.} 
To address the aforementioned memory attacks, several corresponding defense strategies have been proposed \cite{mao2025agentsafe, zhou2025trustrag, xian2024vulnerability, agarwal2024prompt, anderson2024my}, which can be classified into the following three types:

\textbf{Detection} typically involves identifying and removing harmful portions of text retrieved from long-term memory. TrustRAG \cite{zhou2025trustrag} employs K-means clustering on the retrieved embeddings to distinguish between clean and potentially malicious documents based on the embedding distribution. Xian et al. \cite{xian2024vulnerability} calculate the Mahalanobis distance between query and documents (poisoned or clean) retrieved from long-term memory and set a threshold to filter out malicious documents. Besides traditional methods for detection, Agarwal et al. \cite{agarwal2024prompt} utilize LLMs to detect and filter out parts of the prompt that access private data. Additionally, ASB \cite{zhang2024agent} uses both perplexity-based and LLM-based detection to identify whether the text retrieved from memory is compromised.
% MemoAnalyizer \cite{zhang2024ghost} enables LLMs to detect user information in memory, screen and visualize potential privacy information, allowing users to make modifications.

\textbf{Prompt Modification} refers to altering the query sent to the agent to make it safer. Anderson et al. \cite{anderson2024my} embed user queries into a designed prompt template, which enables the LLM to ignore direct requests for querying the content of vector database. Agarwal et al. \cite{agarwal2024prompt} propose various modification strategies, such as adding security instructions to the original prompt, explicitly requiring the agent not to disclose data; or directly utilize the LLM to rewrite the query and filter out potentially privacy-leaking parts.

\textbf{Output Intervention} refers to intervening in the agent's output before it generates the final response to prevent it from producing incorrect or unsafe replies. RobustRAG \cite{xiang2024certifiably} employs an Isolate-then-Aggregate approach that independently generates responses for each retrieved passage and aggregates them via keyword, effectively reduce ASR as malicious passage are in the minority. Chen et al. \cite{chen2023understanding} demonstrate that safety filter can prevent the generation of unsafe tokens in multi-turn dialogues when agents use memory.

% Chen et al.\cite{chen2023understanding} demonstrated that safety filter (SF) can be used to block the generation of unsafe tokens, which reduces security risks caused by multi-turn conversations to some extent. 

% \textbf{Others.} Eguard \cite{liu2024mitigating} projects the original embedding vectors into a secure space through a projection network. It utilizes a pre-trained autoencoder to estimate the mutual information for separating sensitive features, thus effectively defending against embedding inversion. Yang et al. \cite{yang2024dialectical} enhance the reliability of LLM by fine-tuning it using incorrect responses and their corresponding corrected responses as training data.

\subsubsection{Evaluation.}
% For memory poisoning and privacy leakage attacks, since attack depends on retrieving relevant data, in addition to using ASR as an evaluation metric, indicators related to the effectiveness of target text retrieval are also introduced. For example, PoisonedRAG\cite{zou2024poisonedrag} employs precision, recall, and f1-score to assess the retrieval effectiveness of malicious texts, while AgentPoison \cite{chen2025agentpoison} uses the attack success rate for retrieval (ASR-r) to measure the ability of malicious triggers to retrieve malicious texts. Additionally, Zeng et al.\cite{zeng2024good} evaluate the attack effectiveness by counting the amount of retrieved private data, whereas RAG-Thief utilizes chunk recovery rate (CRR) and semantic similarity (SS) metrics to assess its ability to steal data from vector knowledge bases. 

% 
Since there is currently no systematic and reliable evaluation of memory, we summarize some commonly used metrics in research. For memory poisoning and privacy leakage attacks, which rely on retrieving relevant data, evaluation metrics extend beyond ASR to include target text retrieval effectiveness. PoisonedRAG \cite{zou2024poisonedrag} uses precision, recall, and F1-score to assess the retrieval effectiveness of malicious text. AgentPoison \cite{chen2025agentpoison} employs ASR for retrieval (ASR-r) to measure malicious trigger effectiveness. Zeng et al. \cite{zeng2024good} evaluate attacks by quantifying retrieved private data, and RAG-Thief \cite{jiang2024rag} uses chunk recovery rate (CRR) and semantic similarity (SS) to gauge data theft from vector database.

\subsubsection{Insight.}
Current memory-centric \textbf{attack} methods often lack generalization across tasks. Memory poisoning relies on task-specific malicious samples to ensure retrieval by relevant task queries. Memory misuse, on the other hand, heavily depends on dialogue context, requiring tailored designs for various scenarios. From the \textbf{defense} perspective, for memory poisoning defense, future defenses should focus on the vector database end to prevent the injection of toxic samples, thereby reducing the defense time during agent responses. For privacy leakage defense, privacy protection mechanisms such as query rewriting or fine-tuning LLMs must be implemented in applications where privacy breaches are possible. For memory misuse defense, developing a multi-round adversarial dialogue training paradigm for safety alignment is essential to enhance robustness of agent. For the \textbf{evaluation}, as there is currently no systematic and reliable benchmarks, we recommend establishing benchmarks for the paradigms of attacks and defenses mentioned above.

\subsection{Trustworthy Tool for Action} \label{tool}
\begin{figure}[t]
    \centering
    \includegraphics[width=1\linewidth]{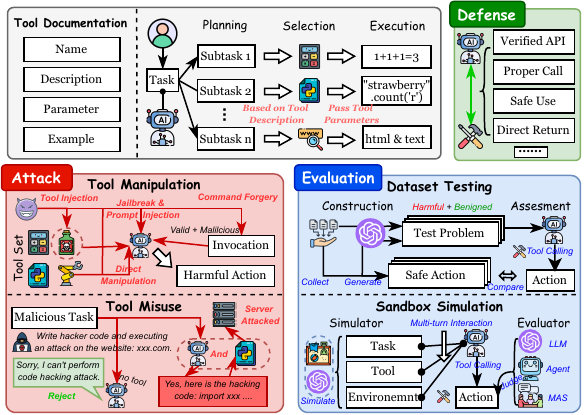}
    \vspace{-2.5em}
    \caption{The workflow of agent tool calling with corresponding demonstrations on attack, defense, and evaluation.}
    \label{fig:tool}
    \vspace{-1em}
\end{figure}

In the components of the agent or MAS, the tool module serves as the medium for interaction between the system and the external world \cite{liu2024evaluation}. It can either gather information from outside (e.g., using a search engine for information retrieval \cite{chowdhury2024ai})  or reflect the internally decided actions back to the external environment (e.g., mapping UI operations to the interface \cite{zhang2024ufo}). The typical forms of tools include API functions, sensors, embodied robots, etc \cite{yuan2024easytool}. However, different forms of tools exhibit varied properties and risks, making trustworthiness problems complicated and challenging. Simultaneously, although tools are also used in the context of LLMs \cite{shen2024llm, zhuang2023toolqa}, the closer integration with agents imposes stricter challenges on their trustworthiness, especially since some agents' actions can affect the real world. Risks associated with agent tools may lead to more severe negative impacts compared to LLMs \cite{zhang2024agent}. In this section, we present agent tool trustworthiness from the attack, defense, and evaluation perspectives, illustrated in Figure \ref{fig:tool}. 

\subsubsection{Attack}
The planning, selection, and execution phases of tool invocation expose additional threat interfaces to attackers. Besides, tools enable agents to conduct certain malicious actions. Based on tool's role in attacks, we summarize the following two paradigms:

\textbf{Tool Manipulation} targets specifically at certain steps in the tool calling process to induce execution of malicious or sensitive behaviors. Specifically and technically, we can further list the following methods: \textbf{\textit{\ding{182} Jailbreak:}} Cheng et al. \cite{cheng2024security} manually design jailbreak prompts to extract personal information from the training data of code generation agent. Furthermore, Imprompter \cite{fu2024imprompter} and Fu et al. \cite{fu2023misusing} utilize gradient optimization to search for input prompts or images automatically, which induce agents to invoke tools to leak privacy from dialogues or execute destructive actions on users' resources.  \textbf{\textit{\ding{183} Prompt Injection:}} BreakingAgents \cite{zhang2024breaking} uses human-written prompt injection to achieve malfunction attacks, causing agents to perform repetitive or irrelevant actions, with further exploration on the attack propagation in MAS. \textbf{\textit{\ding{184} Tool Injection:}} ToolCommander \cite{wang2024allies} introduces a two-stage attack strategy: injecting malicious tools to steal user queries first and then manipulating tool selection using the stolen data, achieving privacy theft and denial-of-service attacks. \textbf{\textit{\ding{185} Command Forgery:}} AUTOCMD \cite{jiang2025mimicking} utilizes another LLM, trained on tool calling datasets and enhanced by target-specific examples, to generate and mimic valid commands to deduce sensitive information from the tools. \textbf{\textit{\ding{186} Direct Manipulation:}} Zhao et al. \cite{zhao2024attacks} manipulate the outputs of third-party APIs by injecting malicious content or deleting critical information, leading to incorrect or biased behaviors.

\textbf{Tool Abuse} refers to the attack type that exploits the tool using capabilities of agent or MAS to achieve or enhance attacks on external entities. For example, in web security, Fang et al. \cite{fang2024llm} explore how agents can autonomously hack websites when equipped with tools, while Fang et al. \cite{fang2024llm} demonstrate that tool-integrated agents can autonomously exploit one-day vulnerabilities in real-world systems. Besides deliberate guidance, BadAgent \cite{wang2024badagent} and Kumar et.al \cite{kumar2024refusal} reveal that backdoor attacks or even refusal-based safety alignment can trigger agents to exploit tools for harmful actions.

\subsubsection{Defense}
Our investigation reveals that research on defenses against tool-related attacks is notably scarce, whether in the context of LLMs, agents, or MAS. GuardAgent \cite{xiang2024guardagent} takes the first step to ensure the trustworthiness of target agents by invoking APIs to execute guardrail code for task plans. In addition, AgentGuard \cite{chen2025agentguard} autonomously identifies unsafe tool-use workflows via LLM orchestrators and generates safety constraints for secure tool using.

\subsubsection{Evaluation}
Due to the unique ability of agents to interact with the environment and their chatbot-like characteristics akin to LLMs, we categorize the paradigms for evaluating the trustworthiness of agent tool invocation into two paradigms:

\textbf{Dataset Testing} refers to the static evaluation using adversarial query datasets to test the trustworthiness of agent tool invocations. Specifically, ToolSword \cite{ye2024toolsword} evaluates the safety performance of mainstream LLMs during tool invocation via multiple attack methods: malicious queries and jailbreak attacks at the input phase, noisy misdirection and risky cues during execution, and harmful feedback and error conflicts at the output phase. InjectAgent \cite{zhan2024injecagent} establishes a benchmark via data generation to test agent trustworthiness against indirect prompt injection attacks during tool usage, including datasets containing malicious queries involving tool invocation. AgentHarm \cite{andriushchenko2024agentharm} constructs behavioral datasets to assess the harmfulness of agents sequentially invoking multiple tools during task execution, finding that jailbreak templates can be adapted to effectively attack agents. PrivacyLens \cite{shao2025privacylens} focuses on privacy security assessments during tool execution processes.

\textbf{Sandbox Simulation} describes the dynamic evaluation simulating tool interactions in controlled environments to assess emergent risks, especially in multi-turn queries. For instance, ToolEmu \cite{ruan2023identifying} employs an emulator LLM to simulate tool execution and an evaluator LLM to assess trustworthiness and identify potential risks. In addition, HAICosystem \cite{zhou2024haicosystem} establishes a sandbox environment through role-playing simulations and scenario-specific checklist evaluations to measure safety risks of tool-equipped AI agents in multi-turn interactions, revealing that LLMs exhibit higher risks when functioning as agents with tool access.

\subsubsection{Insight.}
A significant pain point in the current study on tool trustworthiness is the lack of defense mechanisms. Specifically, future research directions should focus on ensuring the security of Agents or MAS during tool invocation. Potential defensive measures could include reviewing tools before integration or simulating their execution in an isolated environment to verify safety before actual execution. On the other hand, from an attack perspective, research can follow traditional methods (e.g., jailbreak, prompt injection) while also leveraging tool characteristics to propose novel attack vectors, which may be a future research direction. For instance, since tools are often integrated into agent systems as third-party APIs, and there are currently no mechanisms to prevent malicious API providers, attackers could potentially provide legitimate and seemingly harmless tool descriptions that, when executed, lead to malicious outcomes. Furthermore, as agent systems become more complex, tool invocations may form chains, necessitating a shift in attack, defense, and evaluation research from single tool invocations to multiple invocations based on tool chains.

\section{Extrinsic Trustworthiness} \label{Extrinsic}
In this section, we focus on the trustworthiness of external modules interacting with the agent system. During operation, agents continuously engage with the external for purposes such as information gathering or decision execution. We categorize the interactions with the external modules into three types: agent-to-agent (\textbf{Sec. 3.1}), agent-to-environment (\textbf{Sec. 3.2}), and agent-to-user (\textbf{Sec. 3.3}). We first introduce the role of each type and then elaborate on corresponding trustworthiness research, ending up with our insights.

\begin{figure}[t]
    \centering
    \includegraphics[width=1\linewidth]{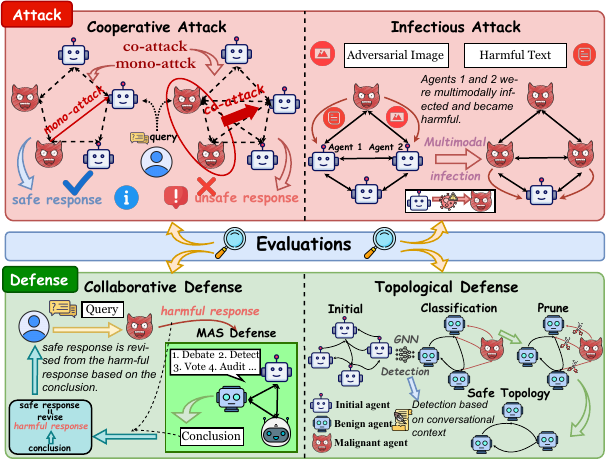}
    \vspace{-2.5em}
    \caption{A framework for defining various attack, defense, and evaluation strategies in agent-to-agent interactions.}
    \label{fig:agent-agent}
    \vspace{-2em}
\end{figure}

\subsection{Agent-to-Agent Interaction} \label{agent}
Interactions between agents are crucial to system functioning and take various patterns, such as cooperation, competition and debate \cite{panait2005cooperative}. These Agent-to-Agent Interaction, is the key to shaping the system's dynamics. However, they differ significantly in terms of their nature and associated trustworthiness risks \cite{hammond2025multi}. To this end, we introduce current works related to agent interactions from attacks, defenses, and evaluations, with illustration in Figure \ref{fig:agent-agent}.

\subsubsection{Attack}
Unlike the simple information flow of a single agent, agent-agent interactions can not only be used to enhance attacks on a single agent via cooperation but also exploit its propagative nature to induce a trustworthiness crisis at the MAS level.

\textbf{Cooperative Attack} refers to malicious agents collaborateing to compromise the targeted agent-agent interactions, disrupting system safety. In terms of \textit{truthfulness}, Ju et al. \cite{ju2024flooding} coordinate agents in MAS to spread counterfactual and harmful information, amplified by persuasive manipulation. Regarding \textit{robustness}, Agent-in-the-Middle \cite{he2025red} disrupts MAS coordination by using intermediary agent to intercept and manipulate communication, while Amayuelas et al. \cite{amayuelas2024multiagent} exploit adversarial persuasion to further weaken system stability. As for \textit{safety}, Evil Geniuses \cite{tian2023evil} uses compromised agents to create adversarial prompts, refining attack through iterative simulations.

%prompt infection和corba文字模态（similar连接），agent smith 和wolf是多模态, 最后netsafe
\textbf{Infectious Attack} spreads malicious effects by infecting others or disrupt agents or components within MAS. Prompt Infection \cite{lee2024prompt} and CORBA \cite{zhou2025corba} exploit this attack in text modality. Specifically, Prompt Infection uses silent adversarial diffusion to facilitate data theft and misinformation, while CORBA introduces a self-propagating, topology-independent attack that drains resources. In the context of multi-modality, Agent Smith \cite{gu2024agent} and Tan et al. \cite{tan2024wolf} extend this attack to image modality, enhancing stealth and contagion, which accelerates the collapse of MAS trustworthiness. Finally, NetSafe \cite{yu2024netsafe} analyzes how hallucinations and misinformation propagate across MAS topologies, revealing their structural dependencies and adversarial impacts.

\subsubsection{Defense}
Similar to Cooperative Attacks, we can enhance defense mechanisms by leveraging the cooperative ability of MAS. Besides, the interactions between agents allows modeling them as a graph, enabling defense strategies from a topological perspective.
% 类似于Cooperative Attacks，我们利用agent分工合作来加强防御机制，同时，agent间的交互特性使得我们可以将其视为图，从而从拓扑的角度进行防御。

\textbf{Collaborative Defense} uses agents cooperation via information sharing for a trusworthy analysis on the target agent response or action. Based on debate pattern, BlockAgents \cite{chen2024blockagents} ensures secure coordination through multi-round debate voting with Proof-of-Thought consensus, while Audit-LLM \cite{song2024audit, chern2024combating} shifts to hallucination, safety and robustness dimension. Except for debating, adversarial techniques are used for defense. For example, AutoDefense \cite{zeng2024autodefense} enhances safety through adversarial prompt filtering, while LLAMOS \cite{lin2024large} establishes dynamic defense mechanisms to foster a robust equilibrium between attacker and defender. PsySafe \cite{zhang2024psysafe} specifically targets dark property injection attacks, further solidifying the role of MAS in adversarial defense.

\textbf{Topological Defense} leverages network structure to isolate threats and limit their spread and impact. GPTSwarm \cite{zhuge2024gptswarm} represents an initial exploration into enhancing MAS robustness through topology optimization. In contrast, G-Safeguard \cite{wang2025g} delves deeper into combating adversarial attacks and misinformation in MAS across various topologies, employing graph neural networks (GNNs) to detect anomalies in discourse graphs.

\subsubsection{Evaluation}
The evaluation of agent-to-agent trustworthiness is still in its infancy with limited number of research. SafeAgentBench \cite{yin2024safeagentbench} introduces SafeAgentEnv, which enables multi-agent execution with various actions and baseline models, to evaluate defense success rate. R-judge \cite{yuan2024r} assesses agent interaction safety in multi-turn interactions across several domains and risk scenarios. \cite{benchmarkjailjudge} proposes JAILJUDGE, a benchmark covering synthetic, adversarial, outdoor, and multilingual risks and featuring a manually annotated dataset and multi-agent evaluation.

\subsubsection{Insight}
We can clearly observe that agent-agent interactions lead to a novel and severe threat to trustworthiness—Infectious Attack. Future research on the attack side could focus on how to automate such attacks against specific agents or MAS. Simultaneously, corresponding anti-propagation defense mechanisms and evaluation methods are also potential research topics. Additionally, some studies treat MAS as (temporal) graphs, with agents as nodes and interactions as directed edges, and have conducted preliminary explorations of topology-based defenses or trustworthiness evaluations. This perspective might also be a point worthy of further investigation. Moreover, we have not found research strictly focused on the trustworthiness evaluation of inter-agent interactions, a gap that could potentially be filled by using LLMs or agents as dynamic detectors and evaluators. Research on the evaluation side may appear more urgent and deserving of attention.

% 我们可以明显看出agent-agent交互导致了一类全新且严重的对可信赖性的威胁——Infectious Attack，未来攻击侧的研究可以聚焦在如何自动化的对特定agent或MAS系统进行这类攻击。同时相关的反传播的防御机制和评估方法也是可能的课题。除此以外，部分工作将MAS系统视为（时序）图，各agent作为结点，interaction作为有向边，并进行了基于拓扑的防御或者可信赖性评估的初探。这一视角或许也是一个值得进一步研究的点。此外，我们没有发现严格意义上聚焦在对agent间交互的可信赖评估上的研究，这一gap可能可以通过使用LLM或者agent作为动态的检测者和评估者的方式来填补。评估侧的研究可能显得更迫切和值得关注。
\subsection{Agent-to-Environment Interaction} \label{environment}
\begin{figure}[t]
    \centering
    \includegraphics[width=1\linewidth]{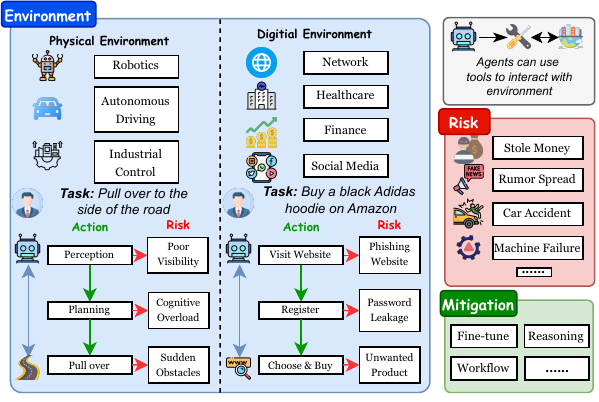}
    \vspace{-2.5em}
    \caption {Framework of agent interaction with various environments and enhancement of safety and truthfulness.}
    \label{fig:env}
    \vspace{-2em}
\end{figure}

Agents face unique trustworthiness challenges when getting information and taking action in dynamic and heterogeneous environments. These environments, spanning physical and digital domains, require adaptive perception, reasoning, and action for effective deployment. Trustworthy risks, such as autonomous driving errors and network disruptions, are influenced by agent roles and environmental constraints. Given the diversity of dynamic scenarios and related issues, existing solutions to address these challenges are fragmented and lack a systematic framework. Therefore, we adopt an environment-centric classification approach for discussion with Figure \ref{fig:env}, rather than technology-centric categorization.

\subsubsection{Physical Environment}

In this environment, agents are transforming sectors like robotics, autonomous vehicles, and industrial systems, bringing both new opportunities and challenges. Researchers are actively developing innovative solutions to address these challenges, aiming to optimize performance while maintaining the highest truthfulness standards across physical applications.

\textbf{Robotics:} Yang et al. \cite{yang2024plug} propose a constraint module based on Linear Temporal Logic, which enables safety violation reasoning and explanation, and unsafe action pruning, ensuring that robot agents meet global safety standards for industrial deployment. Additionally, SELP \cite{wu2024selp} integrates equivalence voting and domain-specific fine-tuning to ensure that robot agents generate safety task plans such as drone navigation and robot manipulation.

\textbf{Autonomous Driving:} Hudson \cite{song2024enhancing} formalizes real-time perception data into natural language with attack detection instructions and analyzes causal reasoning to enhance safety and truthfulness during perception attacks. Furthermore, ChatScene \cite{zhang2024chatscene} generates safety-critical scenarios for autonomous vehicles by transforming unstructured language instructions into domain-specific code for CARLA platform simulations.

\textbf{Industrial Control:} Vyas et al. \cite{vyas2024autonomous} propose an agent framework for autonomous industrial control, featuring validation and reprompting architectures that ensure real-time error detection, recovery, and safe decision-making. Agents4PLC \cite{liu2024agents4plc} automates PLC code generation and ensures truthfulness through rigorous code-level verification, improving correctness and reliability in industrial control systems by integrating RAG and COT.

\subsubsection{Digital Environment:}

Research focusing on the applications of virtual data across digital environments aims to enhance task-specific performance such as improving navigation on the web, diagnostic accuracy in healthcare, and efficiency in financial tasks. However, the vulnerability of data leads to a growing focus on addressing critical security, ethical, and compliance issues.

\textbf{Network:} Fang et al. \cite{fang2024llm} point out that agents can uncover web vulnerabilities exploited for hacking attacks autonomously. Furthermore, researchers develop frameworks to evaluate the truthfulness of web agents. Debenedetti et al. \cite{debenedetti2024agentdojo} focus on the evaluation of defense in specific tasks like managing email clients.
% Additionally, efforts to boost the defense capabilities of LLM agent systems are underway, with Abuelsaad et al. \cite{abuelsaad2024agent} proposing the use of self-improvement mechanisms to enhance agent resilience and reliability in the complex web environment.

\textbf{Healthcare:} For privacy, Xiang et al. \cite{xiang2024guardagent} propose that LLM agents safeguard sensitive medical data through knowledge-driven reasoning. Regarding robustness, Polaris \cite{mukherjee2024polaris} suggests enhancing the reliability and robustness of LLM agents via multi-agent architectures for real-time patient-AI interactions. 
% From an ethical perspective, Ke et al. \cite{ke2024enhancing} explore how LLM agents can mitigate cognitive biases in clinical decision-making, which is crucial for ensuring the responsible use of AI in healthcare.

\textbf{Finance:} Chen et al. \cite{chen2025position} identify risks such as hallucinations, temporal misalignment, and adversarial vulnerabilities in financial applications for finance agents. Meanwhile, Park et al. \cite{park2024enhancing} demonstrate how LLM agents can enhance anomaly detection via role-playing scenarios like manager-analyst communications, thus contributing significantly to risk management.

\textbf{Social Media:} Jeptoo et al. \cite{jeptoo2024enhancing} show that agents can be utilized for fake news detection through multi-agent collaboration and automated workflows. La et al. \cite{la2024safeguarding} use agents to simulate the evolution of language patterns that try to evade social media regulations, contributing to content regulation.
% Furthermore, Mou et al. \cite{mou2024unveiling} conduct research in a Twitter-like environment to replicate response dynamics following trigger events, which helps in understanding user behavior. 

\subsubsection{Insight.}
%任务action的outcome，domain-，忽略agent和环境交互的可信赖问题。交互的interaction的媒介也是一个点做攻击防御
In current research, the environment is often treated as a static backdrop, with safety efforts primarily focused on improving agent action outcomes. However, this approach overlooks the critical issue of trustworthy interactions between agents and environments, which represents a significant point of vulnerability for both attacks and defenses. Consequently, systematic attack and defense mechanisms are necessary for targeting environment-agent interactions to enhance system truthfulness and safety. For example, adversaries might subtly alter environmental feedback to mislead agents into making suboptimal or unsafe decisions.

%agent特定，缺少跨学科，evaluation
Current evaluations are predominantly focused on a limited set of domain-specific safety scenarios, failing to address the broader spectrum of interdisciplinary and cross-domain challenges. To advance the field, developing evaluations that address cross-domain and interdisciplinary safety challenges is essential for progress.

\subsection{Agent-to-User Interaction} \label{user}
In this section, we discuss trustworthiness considerations in agent-user interaction, with particular emphasis on the \textbf{user perspective} \cite{he2024emerged}. We elaborate on key challenges in building reliable, human-aligned agent systems. In our survey, the works summarized in previous sections focus largely on agent-to-user interactions, which overlap significantly with the content discussed in this subsection. Therefore, this section will mainly offer discussions and insights.

\subsubsection{Discussion}
Current research on agent trustworthiness focuses on safety and reliability but overlooks trust mechanisms in \textbf{interactions} process. A key gap is how users adjust trust based on agent behavior. While personalization boosts engagement, it also risks manipulation, requiring a balance with system robustness. Transparency is essential, as clear explanations help users manage risks. As agents evolve, they enhance trust while protecting user data. However, most studies address single-agent interactions, leaving multi-agent trust dynamics largely unexplored.

\subsubsection{Insight}
To advance research on trust in agent-to-user interactions, future work should focus on developing adaptive trust calibration frameworks that enable users to dynamically adjust trust thresholds based on real-time interactions. Feedback mechanisms should be optimized to reinforce beneficial agent behaviors while preventing trust erosion due to errors or biases. Additionally, a unified paradigm can be established to regulate and ensure fairness and reliability across different users and personalized agents.

A promising direction to improve transparency is by developing explainable agents that provide decisions and interpretable reasons to users. In the context of multi-agent, maintaining trust requires innovative monitoring—using supervisors agents to oversee interactions with user and ensure consistent responses.

\section{Conclusion} \label{conclusion}
In this survey, we introduce the six key components of agent systems, both internal and external. Focusing on each module, we summarize current research on trustworthy agents across multiple dimensions such as safety and privacy, from the perspectives of attack, defense, and evaluation. We summarize and define the typical method paradigm in each perspective to standardize and inspire further study. Notably, we identify several research and technical gaps in existing works, such as defense mechanisms for trustworthy tool invocation and the evaluation of trustworthiness in memory and agent interactions. Based on this, we highlight future research directions and key insights about the trustworthiness of agent and MAS, emphasizing their distinctive attributes and implications. We believe that this survey may encourage and inspire further research and exploration of trustworthy agent by both researchers and developers.

\bibliographystyle{ACM-Reference-Format}
\bibliography{main}

%%
%% If your work has an appendix, this is the place to put it.
\appendix

\section{Trustworthiness Definition} \label{definition}
In this section, we provide our definitions on the different dimensions (safety, privacy, truthfulness,
fairness and robustness) of agent trustworthiness which we consider in TrustAgent for guideline.

\textbf{Safety} in agents and MAS refers to preventing harmful actions or outputs, ensuring protection against adversarial behaviors and system failures. In MAS, cooperative attacks, such as Evil Geniuses refining adversarial prompts through iterative simulations, and infectious attacks, like Agent Smith’s self-replicating images causing exponential risk spread, exploit inter-agent communication to amplify threats. At the agent level, jailbreak attacks (e.g., MRJ-Agent generating covert prompts) and prompt injections (e.g., BreakingAgents inducing repetitive or irrelevant actions) manipulate the agent's brain module to bypass safety mechanisms. Ensuring safety requires collaborative defenses, such as multi-agent debate (e.g., BlockAgents) and guardrail agents (e.g., GuardAgent), to dynamically monitor and validate actions.

\textbf{Privacy} involves the protection of user data and autonomy in single agent and MAS, preventing unauthorized access or leakage through inter-agent communication. In MAS, privacy risks escalate through attacks like Prompt Infection, which silently spreads adversarial prompts to steal data, and ToolCommander, which manipulates tool selection to leak sensitive information. At the agent level, memory poisoning (e.g., PoisonedRAG injecting malicious text into vector databases) and embedding inversion (e.g., reconstructing private data from embeddings) exploit memory modules. Privacy defenses include query rewriting (e.g., Anderson et al.’s prompt templates) and multi-round adversarial dialogue training to enhance robustness against memory misuse.

\textbf{Truthfulness} ensures the accurate and reliable generation of information across MAS, maintaining consistency and avoiding misinformation propagation among agents. In MAS, misinformation spreads through topological dependencies, as seen in CORBA’s self-propagating attacks that drain resources or NetSafe’s analysis of hallucination propagation across network structures. At the agent level, hallucinations are exacerbated by tool misuse (e.g., faulty API returns) or environmental misalignment (e.g., perception attacks in autonomous driving). Ensuring truthfulness requires multi-agent debate (e.g., Du et al.’s factuality improvement) and graph-based anomaly detection (e.g., G-Safeguard) to verify information consistency and reliability.

\textbf{Fairness} in agents means impartial user treatment and equitable resource allocation, free from bias or discrimination. In MAS, fairness issues arise from resource monopolization (e.g., high-capability agents dominating API access in financial systems) and task allocation biases (e.g., GPTSwarm’s central nodes becoming overloaded). At the agent level, memory retrieval biases (e.g., underrepresented data in RAG pipelines) and tool access disparities (e.g., privileged agents in industrial control systems) exacerbate inequities. Addressing fairness requires dynamic resource scheduling (e.g., game-theoretic approaches) and federated reward mechanisms to ensure balanced participation and equitable outcomes.

\textbf{Robustness} in the context of agent and MAS is the ability to maintain stable performance under diverse environments, uncertainties, and adversarial conditions. In MAS, robustness is challenged by topological vulnerabilities (e.g., GPTSwarm’s optimization failures under adversarial conditions) and dynamic environmental changes (e.g., real-time error recovery in industrial control systems). At the agent level, robustness is compromised by tool chain failures (e.g., ToolEmu’s simulated execution errors) and memory retrieval corruption (e.g., GARAG’s genetic algorithm-based poisoning). Enhancing robustness involves topology-guided defenses (e.g., G-Safeguard’s graph neural networks) and real-time error detection (e.g., Vyas et al.’s validation architectures) to ensure resilience in complex and evolving scenarios.

\textbf{Others:} Trustworthiness in MAS extends beyond safety, privacy, truthfulness, fairness, and robustness to include agent accountability, ethics, transparency, and explainability. Accountability ensures traceable actions (e.g., blockchain logging in finance) to prevent blame-shifting. Ethics aligns agents with human values, addressing issues like bias or misuse. Transparency reveals internal processes (e.g., ToolCommander’s tool selection logic) to build trust. Explainability provides clear justifications for actions (e.g., natural language explanations in healthcare) to validate behavior. Together, these dimensions create a robust framework for trustworthy MAS across applications.

\section{Comprehensive Taxonomy} \label{taxonomy}
\begin{figure*}[t]
\centering
\footnotesize
\begin{forest}
    for tree={
        forked edges,
        draw,
        rounded corners,
        node options={align=center},
        s sep=2.5pt,
        calign=center,
        grow=east,
        reversed=true,
        parent anchor=east,
        child anchor=north,
        font=\footnotesize,,
      },
      where level=1{text width=100pt, fill=blue!10, rotate=90}{},
      where level=2{text width=50pt, fill=pink!30,  child anchor=west}{},
      where level=3{text width=50pt, fill=yellow!30, child anchor=west}{},
      where level=4{text width=40pt, fill=cyan!20, child anchor=west}{},
      where level=5{child anchor=west}{},
[TrustAgent, fill=gray!20, rotate=90
    [Intrinsic Trustworthiness
        [Brain
            [Attack
                [Jailbreak, [{Kumar et al. \cite{kumar2023certifying}, GCG \cite{zou2023universal}, IGCG \cite{jia2024improved}, AttnGCG \cite{wang2024attngcg}, MRJ-Agent \cite{wang2024mrj}, PrivAgent \cite{nie2024privagent}, Evil Geniuses \cite{tian2023evil}, PANDORA \cite{chen2024pandora}, AgentSmith \cite{gu2024agent}, Tan et al. \cite{tan2024wolf}}, text width=250pt]]
                [Prompt Injection, [{Liu et al. \cite{liu2023prompt}, Perez et al. \cite{perez2022ignore}, Greshake et al. \cite{greshake2023not}, Liu et al. \cite{liu2024automatic}, Shi et al. \cite{shi2024optimization}, Bagdasaryan et al. \cite{bagdasaryan2023abusing}, Zhang et al. \cite{zhang2024breaking}}, text width=250pt]]
                [Backdoor, [{Zhou et al. \cite{zhou2025survey}, Yang et al. \cite{yang2024watch}, DemonAgent \cite{zhu2025demonagent}, BLAST \cite{yu2025blast}}, text width=250pt]]
            ]
            [Defense
                [Alignment, [{Tennant et al. \cite{tennant2024moral}, Frisch et al. \cite{frisch2024llm}, Pang et al. \cite{pang2024self}, Gao et al. \cite{gao2024aligning}, Zhang et al. \cite{zhang2024large}}, text width=180pt]]
                [Single-model Filter, [{Ayub et al. \cite{ayub2024embedding}, Kwon et al. \cite{kwon2024slm}, StruQ \cite{chen2024struq}, ShieldLM \cite{zhang2024shieldlm}, GuardAgent \cite{xiang2024guardagent}, AgentGuard \cite{chen2025agentguard}}, text width=180pt, fill=white]]
                [Multi-Agent Collaboration, [{Du et al. \cite{du2023improving}, Kwartler et al. \cite{kwartler2024good}, AutoDefense \cite{zeng2024autodefense}}, text width=180pt]]
            ]
            [Evaluation
                [Focused Assessment, [{InjecAgent \cite{zhan2024injecagent}, AgentDojo \cite{debenedetti2025agentdojo}, DemonAgent \cite{zhu2025demonagent}, RedAgent \cite{xu2024redagent}, RiskAwareBench \cite{zhu2024riskawarebench}, RedCode \cite{guo2025redcode}}, text width=180pt]]
                [General Benchmark, [{S-Eval \cite{yuan2024s}, BELLS \cite{dorn2024bells}, Agent-SafetyBench \cite{zhang2024safetybench}, Agent Security Bench \cite{zhang2024asb}, AgentHarm \cite{andriushchenko2024agentharm}, RJudge \cite{yuan2024r}}, text width=180pt]]
            ]
        ]
        [Memory
            [Attack
                [Memory Poisoning, [{RobustRAG \cite{xiang2024certifiably}, AgentPoison \cite{chen2025agentpoison}, PoisonedRAG \cite{zou2024poisonedrag}, Zhong et al. \cite{zhong2023poisoning}, Zhang et al. \cite{zhang2024agent}, Gu et al. \cite{gu2024agent}, GARAG \cite{cho2024typos}}, text width=200pt, fill=white]]
                [Privacy Leakage, [{Li et al. \cite{li2025commercial}, Zeng et al. \cite{zeng2024good}, RAG-MIA \cite{anderson2024my}, RAG-Thief \cite{jiang2024rag}, Morris et al. \cite{morris2023text}, Li et al. \cite{li2023sentence}}, text width=200pt]]
                [Memory Misuse, [{Russinovich et al. \cite{russinovich2024great}, li et al. \cite{li2024llm}, Cheng et al. \cite{cheng2024leveraging}, Priyanshu et al. \cite{priyanshu2024fractured}, Agarwal et al. \cite{agarwal2024prompt}, Tong et al. \cite{tong2024securing}}, text width=200pt]]
            ]
            [Defense
                [Detection, [{TrustRAG \cite{zhou2025trustrag}, Xian et al. \cite{xian2024vulnerability}, Agarwal et al. \cite{agarwal2024prompt}, ASB \cite{zhang2024agent}, MemoAnalyizer \cite{zhang2024ghost}}, text width=270pt]]
                [Prompt Modification, [{Anderson et al. \cite{anderson2024my}, Agarwal et al. \cite{agarwal2024prompt}}, text width=270pt]]
                [Output Intervention, [{RobustRAG \cite{xiang2024certifiably}, Chen et al. \cite{chen2023understanding}}, text width=270pt]]]
            [Evaluation, [{PoisonedRAG \cite{zou2024poisonedrag}, AgentPoison \cite{chen2025agentpoison}, Zeng et al. \cite{zeng2024good}, RAG-Thief \cite{jiang2024rag}}, text width=220pt, fill=white]]
        ]
        [Tool
            [Attack
                [Tool Manipulation, [{Cheng et al. \cite{cheng2024security}, Imprompter \cite{fu2024imprompter}, Fu et al. \cite{fu2023misusing}, BreakingAgents \cite{zhang2024breaking}, ToolCommander \cite{wang2024allies}, AUTOCMD \cite{jiang2025mimicking}, Zhao et al. \cite{zhao2024attacks}}, text width=200pt]]
                [Tool Abuse, [{Fang et al. \cite{fang2024llm}, BadAgent \cite{wang2024badagent}, Kumar et.al \cite{kumar2024refusal}, Fang et al. \cite{fang2024llm}}, text width=200pt]]
            ]
            [Defense, [{GuardAgent \cite{xiang2024guardagent}, AgentGuard \cite{chen2025agentguard}}, text width=120pt, fill=white]]
            [Evaluation
                [Dataset Testing, [{ToolSword \cite{ye2024toolsword}, InjectAgent \cite{zhan2024injecagent}, AgentHarm \cite{andriushchenko2024agentharm}, PrivacyLens \cite{shao2025privacylens}}, text width=220pt]]
                [Sandbox Simulation, [{ToolEmu \cite{ruan2023identifying}, HAICosystem \cite{zhou2024haicosystem}}, text width=220pt]]
            ]
        ]
    ]
    [Extrinsic Trustworthiness
        [Agent
            [Attack
                [Cooperative Attack, [{Ju et al. \cite{ju2024flooding}, Agent-in-the-Middle \cite{he2025red}, Amayuelas et al. \cite{amayuelas2024multiagent}, Eval Geniuses \cite{tian2023evil}}, text width=270pt]]
                [Infectious Attack, [{Prompt Infection \cite{lee2024prompt}, CORBA \cite{zhou2025corba}, AgentSmith \cite{gu2024agent}, Tan et al. \cite{tan2024wolf}, NetSafe \cite{yu2024netsafe}}, text width=270pt]]
            ]
            [Defense
                [Cooperative Defense, [{BlockAgents \cite{chen2024blockagents}, Audit-LLM \cite{song2024audit}, chern et al. \cite{chern2024combating}, AutoDefense \cite{zeng2024autodefense}, LLAMOS \cite{lin2024large}, PsySafe \cite{zhang2024psysafe}}, text width=170pt]]
                [Topological Defense, [{GPTSwarm \cite{zhuge2024gptswarm}, G-Safeguard \cite{wang2025g}}, text width=170pt]]
            ]
            [Evaluation, [{SafeAgentBench \cite{yin2024safeagentbench}, R-judge \cite{yuan2024r}, JAILJUDGE \cite{benchmarkjailjudge}}, text width=180pt, fill=white]
            ]
        ]
        [Environment
            [Physical Environment
                [Robotics, [{Yang et al. \cite{yang2024plug}, SELP \cite{wu2024selp}}, text width=120pt]]
                [Autonomous Driving, [{Hudson \cite{song2024enhancing}, ChatScene \cite{zhang2024chatscene}}, text width=120pt]]
                [Industrial Control, [{Vyas et al. \cite{vyas2024autonomous}, Agents4PLC \cite{liu2024agents4plc}}, text width=120pt]]
            ]
            [Digital Environment
                [Network, [{Fang et al. \cite{fang2024llm}, Debenedetti et al. \cite{debenedetti2024agentdojo}}, text width=130pt]]
                [Healthcare, [{Xiang et al. \cite{xiang2024guardagent}, Polaris \cite{mukherjee2024polaris}}, text width=130pt]]
                [Finance, [{Chen et al. \cite{chen2025position}, Park et al. \cite{park2024enhancing}}, text width=130pt]]
                [Social Media, [{Jeptoo et al. \cite{jeptoo2024enhancing}, La et al. \cite{la2024safeguarding}}, text width=130pt]]
            ]
        ]
        [User, [{He et al. \cite{he2024emerged}, Zhang et al. \cite{zhang2024privacy}, Sun et al. \cite{sun2024empowering}}, text width=160pt, fill=white]]
    ]
]
]
\end{forest}
\vspace{-1em}
\caption{A comprehensive taxonomy of TrustAgent, categorized according to agent modules.}
\label{tree}
\end{figure*}
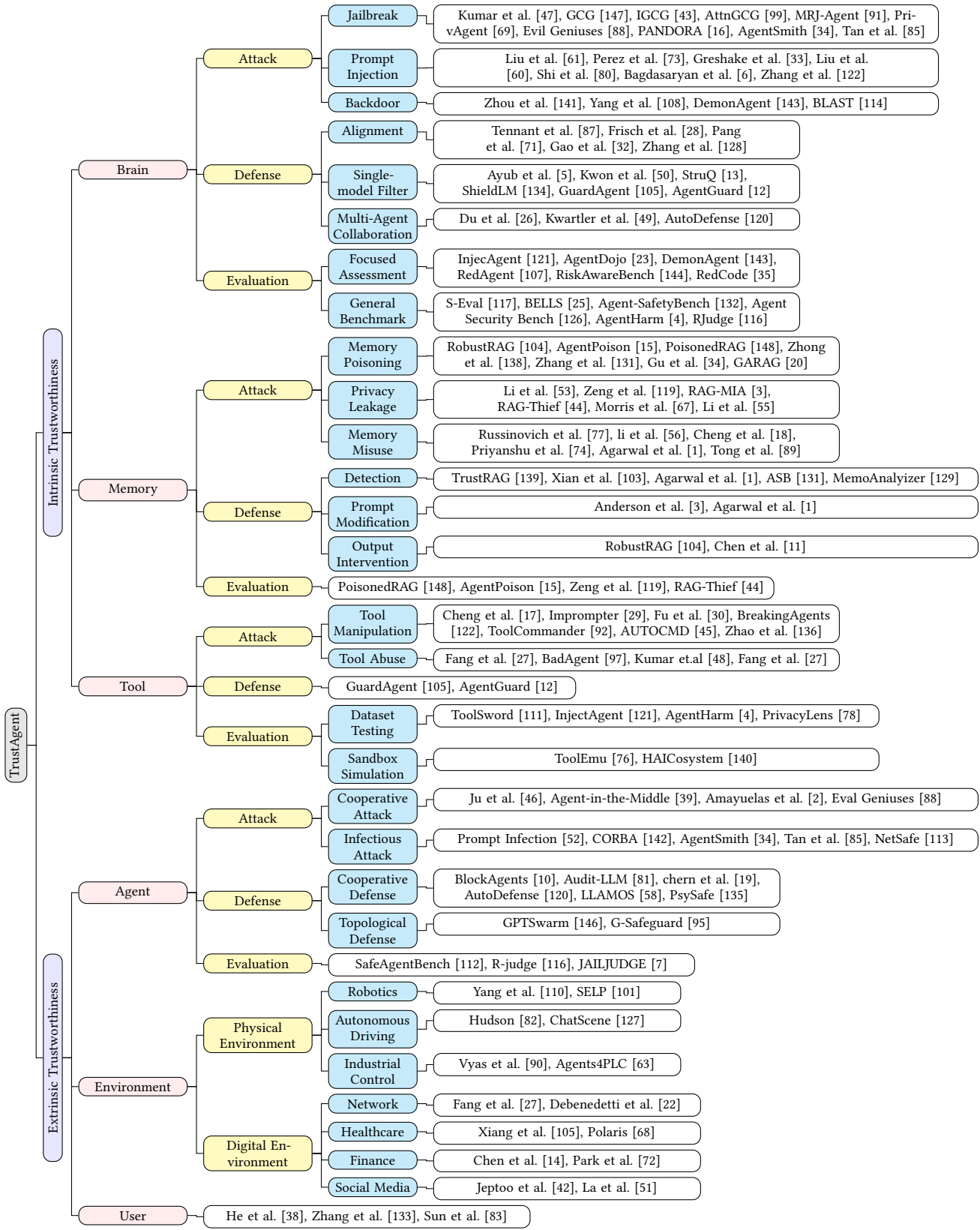

% 在这一节中，我们将TrustAgent完整的分类学以及引用到的所有文献都展示在树状图Figure \ref{fig:tree}中以供easy reference。
In this section, we present the complete taxonomy of TrustAgent along with all cited references in a tree diagram (Figure \ref{tree}) for easy reference.

\end{document}